\documentclass{ws-ijmpa}
\usepackage[super,compress]{cite}
\begin{document}

\markboth{Mar\'ia G\'omez-Rocha}
{Angular momentum decomposition of chiral multiplets in front form}

\catchline{}{}{}{}{}
%
%%%%%%%%%%%%%%%%%%%%%%%%%%%%%%%%%%%%%%%%%%%%%%%%%%%%%%%%%%%%%%%%%%%%

\title{ANGULAR MOMENTUM DECOMPOSITION OF CHIRAL MULTIPLETS IN FRONT FORM}

\author{MAR\'IA G\'OMEZ-ROCHA}

\address{Institut
f\"ur Physik, Universit\"at Graz, A-8010 Graz, Austria
\\
maria.gomez-rocha@uni-graz.at}

\maketitle

\begin{history}
\received{Day Month Year}
\revised{Day Month Year}
\end{history}

\begin{abstract}
In this article we derive the unitary transformation that relates the $q\bar q$ chiral basis
 $\{R; I J^{PC} \}$ 
to the 
$\{I; ^{2S+1}L_J\}$-basis in a front-form framework.
From the most general expression for the 
Clebsch-Gordan coefficients of the Poincar\'e group one can see
 that the chiral limit brings the angular momentum coupling into a simple form that permits the
relation
in terms of $SU(2)$ Clebsch-Gordan coefficients.  
We demonstrate that such a transformation is identical to the one was obtained for canonical 
spin in the instant form.

\end{abstract}

\ccode{PACS numbers:11.30.Rd,11.30.Cp}

\section{Motivation}

It has been shown in Ref. \citen{Glozman:2007at} that there is a unitary transformation that relates 
the $q\bar q$ chiral basis, usually represented as  $\{R; I J^{PC} \}$, 
where $R$ is the index of the chiral representation 
($R=(0,0),\; (1/2,1/2)_a,\; (1/2,1/2)_b,\; \text{or}\; (0,1) + (1,0)$), $I$ 
is the quantum number of isospin, and
$J^{PC}$ indicates the total angular momentum of the state with definite parity and charge,
and the $\{I; ^{2S+1}L_J \}$ basis, which regards the spin-orbit angular momentum coupling used  
in nonrelativistic quantum mechanics.
This allows one to write a particular state belonging to 
a chiral multiplet with quantum numbers $J^{PC}$, as a superposition
of states of the nonrelativistic-inspired
$\{I; ^{2S+1}L_J\}$ classification scheme.

A chiral state with definite parity $|R; I J^{PC}\rangle $ 
can be decomposed  as a superposition of helicity states without definite parity 
$|J\lambda_1\lambda_2\rangle$  
 through  \cite{Glozman:2007ek,Glozman:2007at}
\begin{equation}\label{Leonid1}
|R;IJ^{PC}\rangle =\sum_{\lambda_1 \lambda_2}\sum_{i_1 i_2} 
\chi^{RPI}_{\lambda_1 \lambda_2} C^{Ii}_{(1/2)i_1 (1/2)i_2} 
|i_1\rangle |i_2\rangle |J\lambda_1\lambda_2\rangle
\end{equation}
where $i_{1(2)}$ and $\lambda_{1(2)}$ are individual isospin and  helicity respectively. 
Coefficients $\chi^{RPI}_{\lambda_1 \lambda_2}$ relate the helicity basis to 
the chiral  basis with definite parity in the state. They can be found in Refs.
 \citen{Glozman:2007ek,Glozman:2007at}.
$C^{JM}_{s_1\sigma_1 s_2\sigma_2}$ are the 
usual $SU(2)$ Clebsch-Gordan coefficients.

Two-particle helicity states $ |J\lambda_1\lambda_2\rangle $ can be written in terms of vectors 
in the $\{I; ^{2S+1}L_J\}$ basis once one knows the expression for the matrix elements \cite{Landau}
\begin{equation}\label{Leonid3}
\langle J\lambda_1 \lambda_2| ^{2S+1}L_J\rangle = 
\sqrt{\frac{2L+1}{2J+1}}C^{S\Lambda}_{(1/2)\lambda_1(1/2)-
\lambda_2}C^{J\Lambda}_{L0S\Lambda}
\end{equation}
It represents the angular momentum coupling of a two-particle state with individual
helicities $\lambda_1$,  $\lambda_2$ (with $\Lambda=\lambda_1-\lambda_2$) to 
a system of total spin $S$ and and orbital 
angular momentum $L$. 

Combining (\ref{Leonid1}) and (\ref{Leonid3}) one finds
\begin{eqnarray}\label{chiral:to:LS}
|R;IJ^{PC}\rangle &=& \sum_{LS}\sum_{\lambda_1 \lambda_2}
\sum_{i_1 i_2}\chi^{RPI_{\lambda_1 \lambda_2}}C^{Ii}_{(1/2)i_1(1/2)i_2} 
|i_1\rangle |i_2\rangle \nonumber  \\
&&\times \sqrt{\frac{2L+1}{2J+1}}C^{S\Lambda}_{(1/2)\lambda_1 (1/2)-\lambda_2}
C^{J\Lambda}_{L0S\Lambda}|^{2S+1}L_J\rangle .
\end{eqnarray}
As an example, the $\rho$-like state 
which belongs to the chiral multiplet $|(0,1)+(1,0);11^{--}\rangle$ and 
$|(1/2,1/2)_b ;11^{--}\rangle$ can be represented as \cite{Glozman:2007at}
\begin{eqnarray}
|(0,1)+(1,0);11^{--}\rangle =\sqrt{\frac{2}{3}}|1;^3S_1\rangle +\sqrt{\frac{1}{3}}|1;^3D_1\rangle,\label{rho1} \\
|(1/2,1/2)_b;11^{--}\rangle =\sqrt{\frac{1}{3}}|1;^3S_1\rangle -\sqrt{\frac{2}{3}}|1;^3D_1\rangle.\label{rho2}
\end{eqnarray}

Since both, the chiral and $^{2S+1}L_J$ representations are complete for two-particle 
systems  with
the quantum numbers $I, J^{PC}$, the angular momentum expansion is uniquely determined 
for each
chiral state.
 Chiral symmetry imposes strong restrictions on the spin and angular momentum 
 distribution of  a system. 
The decomposition has been used in Ref. \citen{Glozman:2009rn} to test the chiral symmetry breaking of the
 $\rho$ meson in the infrared, and at the same time, to 
reconstruct its spin and orbital angular momentum content in terms of partial waves.
This was achieved by using interpolators that transform according to 
$|(0,1)+(1,0);11^{--}\rangle$ and $|(1/2,1/2)_b;11^{--}\rangle$.
If chiral symmetry were not broken there would be only two possible chiral states in the meson, while chiral symmetry
breaking would imply a superposition of both. The obtained result in Ref. \citen{Glozman:2009rn}
indicates that the $q\bar q$ component of the $\rho$-meson in the infrared is indeed a superposition of the 
$|(0,1)+(1,0);11^{--}\rangle$ and $|(1/2,1/2)_b;11^{--}\rangle$ chiral states, and therefore chiral 
symmetry turns out to be broken. By using transformation
(\ref{rho1}) and (\ref{rho2}) the partial wave content can be extracted, obtaining for the 
particular case of
 the $\rho$ meson, a nearly pure $^3S_1$ state  \cite{Glozman:2009rn}. 
This is an example of application, see also Refs. \citen{Glozman:2007at,Glozman:2009bt}. 

In this paper we will not discuss  problems in which the chiral basis or its transformation
can play a role as was done in Refs. \citen{Glozman:2009rn,Glozman:2007at} or \citen{Glozman:2009bt}, for instance. 
The problem we want to address here
 is more technical and related to the transformation (\ref{chiral:to:LS})
itself.
The unitary transformation (\ref{chiral:to:LS}) was obtained in the 
instant 
form of relativistic quantum mechanics. 
In this work we
investigate the corresponding expression one should use in the context of 
 approaches that use light-front 
quantization \cite{Brodsky:1997de} or front-form relativistic quantum mechanics \cite{KeisterPolyzou}.
We pose the question whether the transformation  (\ref{chiral:to:LS}) is identical
 in any other form \cite{Dirac:1949cp,KeisterPolyzou}
or if it is a special feature of those that use canonical spin, 
such as the instant- or the point-forms. 
 The problem is not trivial since
in relativistic composite systems the internal degrees of freedom transform among themselves 
nontrivially under rotations \cite{KeisterPolyzou}. Relativity mixes spatial and temporal
components, and as a consequence, one is not allowed to treat boosts and 
angular momentum separately in general. The election of a particular representation matters and in some cases
some of the symmetry properties of the Poincar\'e group  
might not be manifest. 
The front form is of special interest, since rotations do not form a subgroup of the kinematical 
group
and hence, rotational invariance is not manifest. 
On the other hand, front-form boosts form a subgroup 
of the Poincar\'e group, 
and as a result, the front-form Wigner rotation becomes the identity \cite{KeisterPolyzou}. 

In this article we will  show that the unitary transformation derived in Ref. \citen{Glozman:2007at}
in instant form
is indeed identical in the front form of relativistic quantum mechanics. 
The argument resides in the fact that the \textit{generalized
Melosh rotation} that transforms front-form spins to helicity ones, becomes the identity 
when the mass goes to zero \cite{Diehl:2003ny,Soper:1972xc,Lepage:1980fj}. 

\section{Instant-form decomposition}

Due to rotational and translational invariance in 
nonrelativistic quantum mechanics,
the angular momentum coupling of two particles with individual spin and 
orbital angular momentum $(\mathbf s_1,\mathbf l_1)$ and
$(\mathbf s_2,\mathbf l_2)$ to give a composite system of total spin and 
orbital angular momentum $(\mathbf S,\mathbf L)$ is easily realized by using 
the $SU(2)$ Clebsch-Gordan coefficients.
Relativity involves however, a change of representation 
in which the single-particle momenta and spins are replaced by an overall
system momentum and internal angular momentum \cite{KeisterPolyzou}. It is customary to 
use the Clebsch-Gordan coefficients of the Poincar\'e group \cite{KeisterPolyzou}.

The kind  of spin vector can be fully determined through 
the choice of a certain type of boost. 
Canonical boosts are rotationless.
Spin vectors defined through canonical boosts have the advantage 
that in the center of momentum frame they transform under rotations
 in the same way as in nonrelativistic quantum mechanics, and therefore, for a composite system 
one can find a direct decomposition in terms of $SU(2)$ Clebsch-Gordan coefficients. 
The reason is that 
in the canonical case
the Wigner rotation associated with a pure rotation, turns out
to be the rotation itself \cite{KeisterPolyzou}.
This does not hold in general. In the front form, 
rotational invariance is not manifest, and an angular momentum decomposition in terms of
Clebsch-Gordan coefficients requires additional transformations.

Expression (\ref{Leonid3}) can be achieved in a straightforward manner in 
non relativistic quantum mechanics, as well as in the instant form of dynamics or
in any other form that uses canonical spin. 
The derivation of (\ref{Leonid3}) can be found in Ref. \citen{Landau}.
 We will reproduce it here in a basis of eigenstates of 
 the Poincar\'e group
 in order to be able to refer some steps that concern what will be exposed in the next
section. 
We decompose the spin part of a two-particle state with total 
canonical angular momentum $J$ and $\hat z$-component $M$, orbital angular momentum 
$L$ and total spin $s$, in terms of quantum numbers of the constituents in the center of
momentum frame ($\mathbf P = \mathbf 0$), where the relative momentum is expressed as $\mathbf k=\mathbf k_1=-\mathbf k_2$,

\begin{equation}\label{LandauFormel}
|[LS]|\mathbf k| J; \mathbf 0\, M\rangle =
\sum_{M_LM_S}\sum_{\sigma_1\sigma_2}\int d\hat{\mathbf k}\,|\mathbf k \sigma_1 -\mathbf k\sigma_2\rangle 
 C^{SM_S}_{s_1\sigma_1 s_2\sigma_2}
 Y_{LM_L}(\mathbf{\hat{\mathbf k}})\; C^{JM}_{LM_LSM_S},
\end{equation}
where $|\mathbf k \sigma_1 -\mathbf k\sigma_2\rangle  := |\mathbf k \sigma_1\rangle |-\mathbf k\sigma_2\rangle$, 
$s_{1(2)}$ and $\sigma_{1(2)}$ are respectively the individual 
canonical spins and their $\hat z$-projections. 

Given a particular direction $\hat{\mathbf n}$, where the tensor product state can be written
\begin{equation}
 \langle \mathbf{ \hat n}|\mathbf k \sigma_1 -\mathbf k\sigma_2\rangle :=\psi_{s_1\sigma_1}(\mathbf k) 
\psi_{s_2\sigma_2}(-\mathbf k)\delta^2(\mathbf{\hat k -\hat n}),
\end{equation}
one can write
\begin{eqnarray}
 \psi_{JLSM}(\mathbf k)&:=&\langle \mathbf{ \hat n}|[LS]|\mathbf k| J; \mathbf 0\, M\rangle \nonumber\\
&=&\sum_{M_LM_S}\sum_{\sigma_1\sigma_2}\psi_{s_1\sigma_1}(\mathbf k) 
\psi_{s_2\sigma_2}(-\mathbf k)
 C^{SM_S}_{s_1\sigma_1 s_2\sigma_2} Y_{LM_L}(\mathbf{\hat{\mathbf k}})\; C^{JM}_{LM_LSM_S}.
\end{eqnarray} 
In order to express it in terms of 
helicities one needs to transform states with canonical spin to a basis of states with 
helicity spin. 
The unitary transformation that provides this is a Wigner rotation 
whose argument corresponds to the angle between the $z$-axis and the direction of motion 
$\hat{\mathbf k}:=\mathbf k/|\mathbf k|$
 \begin{eqnarray}\label{c:to:h1}
 &&\psi_{s_1\sigma_1}(\mathbf k)=\sum_{\lambda_1}D^{(s_1)}_{\lambda_1\sigma_1}(\hat{\mathbf k})\psi_{s_1 \lambda_1}(\mathbf k),\\
 &&\psi_{s_2\sigma_2}(-\mathbf k)=\sum_{\lambda_2}D^{(s_2)}_{-\lambda_2\sigma_2}(\hat{\mathbf k})\psi_{s_2 -\lambda_2}(\mathbf k).
 \end{eqnarray}
Replacing in (\ref{LandauFormel}) one gets
  \begin{eqnarray}\label{helicity:decomp}
  \psi_{JLSM}(\mathbf k)
  &=&\sum_{M_SM_L} \sum_{\sigma_1\sigma_2}\sum_{\lambda_1\lambda_2}D^{(s_1)}_{\lambda_1\sigma_1}(\hat{\mathbf k})\psi_{s_1 \lambda_1} (\mathbf k)
  D^{(s_2)}_{-\lambda_2\sigma_2}(\hat{\mathbf k})\psi_{s_2 -\lambda_2}(\mathbf k)\; \nonumber\\
  &&\times Y_{LM_L}(\hat{\mathbf k})\;  
  C^{SM_S}_{s1\sigma_1 s2\sigma_2}\; C^{JM}_{LM_LSM_S}.
 \end{eqnarray}

It is convenient to write the spherical harmonics in 
terms of Wigner $D$-functions\footnote{Our notation differs from Ref. \citen{LandauQM} 
by a factor $i^L$ in the definition of the 
phase of the spherical function.}
\begin{equation}
Y_{LM_L}(\hat{\mathbf k})= \sqrt{\frac{2L+1}{4\pi}}D^{L}_{0M_L}(\hat{\mathbf k})
\end{equation}
in such a way that one can make use of the relation for the product of Wigner $D$-functions 
with the same argument for axially symmetric systems \cite{LandauQM}, 
\begin{equation}\label{D:same:arguments}
 D^{(j_1)}_{m_1'm_1}(\mathbf{\hat{w}})D^{(j_2)}_{m_2'm_2}(\mathbf{\hat{w}}) 
=\sum_{j} C^{jm'}_{j_1m_1'j_2 m_2'}D^{(j)}_{mm'}(\mathbf{\hat{w}}) C^{jm}_{j_1m_1j_2 m_2}, 
\end{equation}
with $m=m_1+m_2$,  $m'=m_1'+m_2'$, and $\mathbf{\hat{w}}$ accounting for the Euler angles.
This leads to 
\begin{eqnarray}\label{total:expansion}
\psi_{JLSM}(\mathbf k)&=&\sum_{\lambda_1 \lambda_2}  \sqrt{\frac{2J+1}{4\pi}}D^{J}_{\Lambda M_J}
 (\hat{\mathbf k})\psi_{s_1\lambda_1}(\mathbf k) \psi_{s_2-\lambda_2}(\mathbf k)\nonumber \\
&&\times \sqrt{\frac{2L+1}{2J+1}}C^{S\Lambda}_{s_1 \lambda_1 s_2 -\lambda_2} 
C^{J\Lambda}_{L0S\Lambda}.
\end{eqnarray}

The fact that the Wigner $D$-functions in (\ref{helicity:decomp}) have the same argument, 
is a particular feature of the instant form, and it
is restricted to the rest frame \cite{KeisterPolyzou}.

It is easy to identify the needed matrix elements as
  \begin{equation}\label{sinCG}
\psi_{JLSM}(\mathbf k)=\sum_{\lambda_1 \lambda_2}\psi_{JM\lambda_1\lambda_2}(\mathbf k)\langle JM\lambda_1\lambda_2 
  |^{2S+1}L_JM\rangle,
  \end{equation}
with
 \begin{equation}
  \psi_{JM\lambda_1\lambda_2} (\mathbf k):=  \sqrt{\frac{2J+1}{4\pi}}D^{J}_{\Lambda M_J} (\hat{\mathbf k})\psi_{s_1 \lambda_1}(\mathbf k)\psi_{s_2 -\lambda_2}(\mathbf k)
 \end{equation}
and 
 \begin{equation}\label{final}
 \langle JM\lambda_1\lambda_2 |^{2S+1}L_JM\rangle =
 \sqrt{\frac{2L+1}{2J+1}}C^{S\Lambda}_{s_1 \lambda_1 s_2 -\lambda_2} C^{J\Lambda}_{L0S\Lambda}.
 \end{equation}
This permits the translation from two-particle helicity states with total angular momentum $J$, 
to a one-particle
state of overall orbital angular momentum $L$ and intrinsic spin $S$. The connexion with chirality
is immediately given by (\ref{Leonid1}).
\section{Front-Form Decomposition}
Eq. (\ref{LandauFormel}) describes the angular momentum decomposition of a representation of 
canonical spin into a superposition of representations with canonical spin. 
Because on the front-form rotations do not form a subgroup of the kinematical group of
the Poincar\'e group, 
 the decomposition (\ref{LandauFormel}) is not feasible \textit{a priori}.
In order to analyze the coupling of two representations with individual spin to a 
superposition of representations with total spin for an arbitrary case in relativistic 
quantum mechanics, it is necessary to use a consistent expression of the Clebsch-Gordan 
coefficients of the Poincar\'e group \cite{KeisterPolyzou}.
Front-form angular momentum coupling is well known and it has been widely applied 
to hadron problems in front-form relativistic quantum mechanics. A relation of the type of 
(\ref{chiral:to:LS}), however, has not been established yet in the front form. 
This is the aim of the present 
section.

In the following we will use the normalization criteria and notation of  Ref. \citen{KeisterPolyzou}.
The light-front 
components of the four-momentum are defined by 
$\tilde{\textbf p}:=(p^+=p^0+p^3,\mathbf p_\perp=(p^1,p^2))$, $p^-=p^0-p^3$.
$|  \tilde{\mathbf p} \mu \rangle_f$ represents a single particle state in a front-form 
basis (denoted by $f$), with $\hat z$-spin projection $\mu$. 
The expression 
for the Clebsch-Gordan coefficients of the Poincar\'e group in the front form 
for an arbitrary frame is given by \cite{KeisterPolyzou}
\begin{eqnarray}\label{ff:ClebschGordan}
&&  _f\langle \mathbf{\tilde p}_1 \mu_1 \mathbf {\tilde p}_2 \mu_2 | \,[LS]\,|\mathbf k| J; 
\mathbf {\tilde P} \,M\rangle_f 
 \nonumber\\
&=&\delta(\mathbf {\tilde P} - \mathbf {\tilde p}_1- \mathbf {\tilde p}_2 )\; 
\frac{1}{|\mathbf k|^2}
\delta(\mathbf k (\mathbf{\tilde p}_1,\mathbf {\tilde p}_2)-\mathbf k)
\left|\frac{\partial (\mathbf{\tilde P},\mathbf{k})}{\partial(\mathbf{\tilde p}_1,\mathbf{\tilde p}_2)}\right|^{1/2}\nonumber \\
&&\times \sum_{\sigma_1\sigma_2} D^{(s_1)}_{\mu_1\sigma_1} 
[\text R_{fc}(\mathbf k, \text{m}_1)]D^{(s_2)}_{\mu_2\sigma_2} [\text R_{fc}(-\mathbf k,\text{m}_2)]\nonumber \\
&&\times Y^L_{M_L}(\hat{\mathbf k})C^{SM_S}_{s_1\sigma_1 s_2 \sigma_2} C^{JM}_{LM_L SM_S},
\end{eqnarray}
where
$_f\langle \mathbf{\tilde p}_1 \mu_1 \mathbf {\tilde p}_2 \mu_2|$ 
represents a tensor-product state of two particles with individual momenta 
$\mathbf{\tilde p}_1$ and $\mathbf{\tilde p}_2$ and spin $\hat z$-projections $\mu_1$ and $\mu_2$ respectively. 
The system of two particles moves with a total light-front momentum $\mathbf{\tilde P}$ 
and the individual spins couple to give a total angular momentum $J$ with orbital and 
spin contributions $[LS]$ in the rest frame in the canonical form, and total angular momentum projection 
on the $\hat z$-direction, $M$. Finally, 
 $\mathbf k=\mathbf k_1=-\mathbf k_2$
is used to denote the individual momenta in the rest frame in the canonical form, 
and $\text{m}_1$ and $\text{m}_2$ denote the individual constituent masses (they should not 
be confused with the spin projections, which appear in italics in equation (\ref{D:same:arguments})). 
The arguments of the Wigner $D$-functions are Melosh rotations
which transform states with canonical spin to states with front-form spin and vice versa. Note that
 the rotation depends on the mass in general, producing a different effect on each constituent. Unless
we are dealing with a system of identical constituent masses (e.g. the chiral case),
we will not be able to use the properties of the $D$-function with the same
argument as was done in the instant form. 

The Clebsch-Gordan coefficient (\ref{ff:ClebschGordan}) is consistent with the normalization 
condition for single states
\begin{equation}
 _f \langle \tilde{\mathbf p}' \mu' |  \tilde{\mathbf p} \mu \rangle_f =
\delta(\tilde{\textbf p}-\tilde{\textbf p}')\,\delta_{\mu\mu'}
\end{equation}
and for state vectors of overall momentum $\tilde{\mathbf P}$
\begin{eqnarray}
&& _f \langle [L'S']\,|\mathbf k'|J'; \mathbf {\tilde P}' \,M' |\, [LS]\,|\mathbf k|J; \mathbf {\tilde P} \,M\rangle_f \nonumber\\
 &&\quad \quad= \delta_{M'M}\delta_{j'j}\delta_{l'l}\delta_{s's}\delta(P'^+-P^+) \delta^2(\mathbf P'_\perp-\mathbf P_\perp)\frac{1}{|\mathbf k|^2}\delta(\mathbf k-\mathbf k').
\end{eqnarray}

The problem now is to couple a state of total front-form angular momentum $J$ 
and spin projection $M$, $| [LS]|\mathbf k|, J; \mathbf {\tilde P} \,M\rangle_f $, 
to a tensor-product state of two particles with individual spins described in 
terms of helicities $ _h\langle \mathbf{ p}_1 \lambda_1 \mathbf { p}_2 \lambda_2 |$.

Irreducible representations with different types of spin are related 
to each other through a unitary transformation \cite{KeisterPolyzou}.
The unitary transformation that relates helicity spin to front-form spin  becomes the identity
for massless particles \cite{Diehl:2003ny,Soper:1972xc,Lepage:1980fj}. This means:
\begin{equation}\label{h:to:f}
  |\mathbf{\tilde p}_1 \mu_1 \mathbf{\tilde p}_2\mu_2\rangle_f\stackrel{\text{m}\to 0} {\longrightarrow}\sum_{\lambda_1\lambda_2} 
 |\mathbf{\tilde p}_1 \lambda_1 \mathbf{\tilde p}_2\lambda_2\rangle_h \delta_{\lambda_1\mu_1}\delta_{\lambda_2\mu_2}
\end{equation}
where the subindex $h$  labels helicity states. Front-form spins and 
helicity spins coincide in the chiral limit, and
one is allowed to make use of them without distinction. 
Replacing (\ref{h:to:f}) in (\ref{ff:ClebschGordan}), one obtains the Clebsch-Gordan coefficient
that couples two-particle helicity states to an overall state of the front-form basis, 
\begin{eqnarray}\label{ff:helicity:final}
&&  _h\langle \tilde{\mathbf{ p}}_1 \lambda_1 \tilde{\mathbf { p}}_2 \lambda_2 |\, [LS]\,|\mathbf k |J;
 \mathbf {\tilde P} \,M\rangle_f \nonumber\\
& =&\delta(\mathbf {\tilde P} - \mathbf {\tilde p}_1- \mathbf {\tilde p}_2 )\; 
\frac{1}{|\mathbf k|^2}\;
\delta(\mathbf k (\mathbf{\tilde p}_1,\mathbf {\tilde p}_2)-\mathbf k) \left|\frac{\partial (\mathbf{\tilde P},\mathbf{k})}{\partial(\mathbf{\tilde p}_1,\mathbf{\tilde p}_2)}\right|^{1/2}\nonumber \\
&&\times \sum_{\sigma_1\sigma_2} D^{(s_1)}_{\lambda_1\sigma_1} [\text R_{hc}(\hat{\mathbf k})]D^{(s_2)}_{\lambda_2\sigma_2} [\text R_{hc}(-\hat{\mathbf k})]\nonumber \\
&&\times Y^L_{M_L}(\hat{\mathbf k})C^{SM_S}_{s_1\sigma_1 s_2\sigma_2}C^{JM}_{LM_LS M_S}. 
\end{eqnarray}
Now the Melosh rotations  
$D^{(s_1)}_{\lambda_1\sigma_1} [\text R_{hc}(\hat{\mathbf k})]$ and $D^{(s_2)}_{\lambda_2\sigma_2} 
[\text R_{hc}(-\hat{\mathbf k})]$ are equivalent to the Wigner rotations and they only depend on the direction of $\mathbf k$.
They have exactly the same significance as 
in (\ref{helicity:decomp}): they transform canonical spins into helicity spins.
We are in the position to write the expression for the state in which we are interested
\begin{eqnarray}\label{state:ff}
 | \,[LS]\,|\mathbf k| J;
 \mathbf {\tilde P} \,M\rangle_f 
&=&\sum_{\lambda_1\lambda_2} \int d^3 \tilde p_1 d^3 \tilde p_2 
| \mathbf{\tilde p}_1 \lambda_1 \mathbf {\tilde p}_2 \lambda_2 \rangle_{h} \\
&&\times _h\langle \mathbf{\tilde p}_1 \lambda_1 \mathbf {\tilde p}_2 \lambda_2 |\, [LS]\,|\mathbf k|,J;
 \mathbf {\tilde P} \,M\rangle_f ,\nonumber
\end{eqnarray}
where 
$\mathbf{1}=\sum\int d^3\tilde p_1 d^3\tilde p_2| \mathbf{\tilde p}_1 \lambda_1 
\mathbf {\tilde p}_2 \lambda_2 \rangle_{h} \, _h\langle \mathbf{\tilde p}_1 \lambda_1 \mathbf {\tilde p}_2 \lambda_2 |$
has been introduced.

Reexpresing it in terms of $\mathbf{\tilde P}$ and $\mathbf k$ and
 setting the center of momentum frame, 
$\mathbf{\tilde P}=\mathbf{\tilde 0}:=
(2p^0,0,0,0)$, 
\begin{eqnarray}
| \,[LS]\,|\mathbf k|J;
 \mathbf {\tilde 0} \,M\rangle_f 
&=&\sum_{\lambda_1\lambda_2} \sum_{\sigma_1\sigma_2} \int d\hat{\mathbf k} |\mathbf k\lambda_1 -\mathbf k \lambda_2\rangle  \nonumber\\
&&\times D^{(s_1)}_{\lambda_1\sigma_1} [\text R_{hc}(\hat{\mathbf k})]D^{(s_2)}_{\lambda_2\sigma_2} [\text R_{hc}(-\hat{\mathbf k})]\nonumber \\
&&\times Y^L_{M_L}(\hat{\mathbf k})C^{SM_S}_{s_1\sigma_1 s_2\sigma_2}C^{JM}_{LM_LS M_S}. 
\end{eqnarray}
Setting now a particular direction of motion $\mathbf{ \hat n}$, the integral over 
$d\hat{\mathbf k}$ can be carried out by means of 
\begin{equation}
 \langle\mathbf{ \hat n} |\mathbf k\lambda_1 -\mathbf k \lambda_2\rangle:= 
\psi_{s_1\lambda_1} (\mathbf k)\psi_{s_2\lambda_2} (-\mathbf k) 
\delta(\hat{\mathbf k}-\hat{\mathbf n}).
\end{equation}
And we have
\begin{eqnarray}\label{ff:JLMS}
 \psi_{JLSM}(\mathbf k)&:=& \langle \mathbf n|\,[LS]\,|\mathbf k|J;
 \mathbf {\tilde 0} \,M\rangle_f \nonumber\\
&=&\sum_{\lambda_1\lambda_2} \sum_{\sigma_1\sigma_2} \psi_{s_1\lambda_1} (\mathbf k)\psi_{s_2-\lambda_2} (\mathbf k)  D^{(s_1)}_{\lambda_1\sigma_1} [\text R_{hc}(\hat{\mathbf k})]D^{(s_2)}_{-\lambda_2\sigma_2} [\text R_{hc}(\hat{\mathbf k})]\nonumber \\
&&\times Y^L_{M_L}(\hat{\mathbf k})C^{SM_S}_{s_1\sigma_1 s_2\sigma_2}C^{JM}_{LM_LS M_S}. 
\end{eqnarray}

Proceeding in the same way as in the previous section in the combination of 
the spherical harmonic and the Wigner 
$D$-functions one obtains

\begin{equation}
\psi_{JLSM}(\mathbf{k})=\sum_{\lambda_1 \lambda_2}\psi_{JM\lambda_1\lambda_2}(\mathbf{k})\langle JM\lambda_1\lambda_2 
|^{2S+1}L_JM\rangle
\end{equation}
being again
\begin{equation}\label{ff:sinCG}
 \psi_{JM\lambda_1\lambda_2} (\mathbf{k}):=  \sqrt{\frac{2J+1}{4\pi}}D^{J}_{\Lambda M_J} 
(\hat{\mathbf k})\psi_{s_1\lambda_1} (\mathbf k)\psi_{s_2-\lambda_2} (\mathbf k),
\end{equation}
and 
\begin{equation}\label{repeated:final}
\langle JM\lambda_1\lambda_2 |^{2S+1}L_JM\rangle =
\sqrt{\frac{2L+1}{2J+1}}C^{S\Lambda}_{s_1 \lambda_1 s_2 -\lambda_2} C^{J\Lambda}_{L0S\Lambda}.
\end{equation}

Having found (\ref{repeated:final}), the validity of decomposition
(\ref{chiral:to:LS}) is demonstrated.

Unlike in the instant form, the combination 
of the Wigner $D$-functions 
would not have been possible if we had 
considered particles 
of different masses. Only in the chiral limit, 
or for equal masses,
the eigenstates in the rest frame transform in the same way as 
in nonrelativistic quantum mechanics. 
Note that in general the coupling (\ref{ff:ClebschGordan}) 
involves rotations that depend on the masses, namely 
$D^{(s_1)}_{\mu_1\sigma_1} [\text R_{fc}(\mathbf k, \text{m}_1)]$ and 
$D^{(s_2)}_{\mu_2\sigma_2}[\text R_{fc}(-\mathbf k,\text{m}_2)]$. 
This would have prevented the application of (\ref{D:same:arguments}), 
since the $D$-functions would not have the same arguments, 
and the dependence on the masses would have entered the decomposition, making it
impossible to write (\ref{ff:JLMS}) in the form of a product of (\ref{ff:sinCG}) 
and (\ref{repeated:final}). Moreover, a further rotation would have been necessary in order to 
transform front-from spins to helicity spins, which in the chiral limit 
turns out to be trivial by means of (\ref{h:to:f}).

The result is that decomposition (\ref{chiral:to:LS}) can be therefore 
used to expand chiral states as a
superposition of vectors of the $\{I;^{2S+1}L_J\}$-basis within a front-form framework. 
It can be expressed as
\begin{eqnarray}\label{ffchiral:to:LS}
 |R; IJ^{PC}\rangle_f 
&=& \sum_{LS}\sum_{\lambda_1 \lambda_2}
\sum_{i_1 i_2}\chi^{RPI}_{\lambda_1 \lambda_2}C^{Ii}_{(1/2)i_1(1/2)i_2} 
|i_1\rangle |i_2\rangle \nonumber  \\
&&\times \sqrt{\frac{2L+1}{2J+1}}C^{S\Lambda}_{(1/2)\lambda_1 (1/2)-\lambda_2}
C^{J\Lambda}_{L0S\Lambda}|^{2S+1}L_J\rangle_f.
\end{eqnarray}

\section{Summary}
We have derived the unitary transformation that relates the $q\bar q$ chiral basis to the 
$\{I; ^{2S+1}L_J\}$-basis in a front-form framework. The result turns out to be the same as
in instant form.\cite{Glozman:2007at}

Spin vectors belonging to different representations can be related through a unitary transformation 
\cite{KeisterPolyzou}. 
We have used the feature of the generalized Melosh rotation 
that relates helicity and 
front-form spins, which becomes the identity for massless particles. 
This has made it possible to find a simple
expression in terms of $SU(2)$ Clebsch-Gordan coefficients, by starting from the most general case of 
the Clebsch-Gordan coefficients of the Poincar\'e group. 
The limit $\text{m}\to 0$ eliminates the mass-dependence in the Wigner $D$-functions making it possible to
express the product of $D$-functions with the same argument through a Clebsch-Gordan series for 
axially symmetric systems.

To better understand the significance of this calculation let us recall that
the purpose of the Clebsch-Gordan coefficients of the Poincar\'e group 
is to convert any kind
of spin to canonical spin in the rest frame, in such a way that they can be
added using $SU(2)$ Clebsch-Gordan coefficients \cite{KeisterPolyzou}. 
In the system rest frame, the only feature that distinguishes front-form spins 
from canonical ones is the fact that 
front-form spins are characterized for being invariant under front-form boosts. This difference
is accounted by the Melosh rotation (cf. (\ref{ff:ClebschGordan})).

As a last remark, let us also mention that it would have been possible to develop such a 
decomposition for any type of spin. The Clebsch-Gordan
coefficients of  the Poincar\'e group for an arbitrary form are given in Ref. \citen{KeisterPolyzou}.
Proceeding in an analogous way as before, it is possible to see 
that again the Wigner $D$-functions do not have the same argument, and it 
is not possible to bring them together to an overall rotation by means
of $SU(2)$ Clebsch-Gordan coefficients. 
Only in the chiral limit the rotations are again the same. 
But unlike in the front form, an additional transformation on such arbitrary spins  
into helicity spins is necessary to make the relation to chirality possible.

\section*{Acknowledgments}

I want to thank W. Schweiger and L. Ya. Glozman for many helpful ideas and discussions as well as
for their help in the elaboration of this paper.
I acknowledge the support of the \lq\lq Fond zur
F\"orderung der wissenschaftlichen Forschung in \"Osterreich\rq\rq
(FWF DK W1203-N16).

%\begin{thebibliography}{000} %for 3 digits
%\begin{thebibliography}{00}  %for 2 digits

\end{document}